\documentclass[10pt,a4]{article}

\usepackage{}

\begin{document}
\begin{titlepage}
\title{\large \bf Existence of Time Operator for a Singular  Harmonic Oscillator}
\author{{\normalsize \bf   V. Mikuta-Martinis and M.Martinis}\\
Theoretical Physics Division, Rudjer Bo\v skovi\'c
Institute\\10002 Zagreb, Croatia\\e-mail:vmikuta@irb.hr}

\date{21. March 2005. }
\maketitle
\begin{abstract}
The time operator for a quantum singular  oscillator of the
Calogero-Sutherland type  is constructed in terms of the
generators of the SU(1,1) group. In the space spanned by the
eigenstates of the Hamiltonian, the time operator is not
self-adjoint.  We show, that the time-energy uncertainty relation
can be given the meaning within the Barut-Girardello coherent
states defined  for the singular oscillator.We have also shown the
relationship with the time-of-arrival operator of Aharonov and
Bohm.
\end{abstract}
\vspace{1cm}

\end{titlepage}

\section{Introduction}
The existence of a self-adjoint time operator conjugate to a given
Hamiltonian is a longstanding problem of quantum mechanics. The
unequal role played by time as an observable in classical and
quantum mechanics is the main source of controversy . The problem
arises because we expect observables to be represented in quantum
mechanics by self-adjoint operators.[1,2]. In an attempt to
promote time to be an observable, we have to face  a well-known
argument of Pauli [3]  that such an operator cannot be
self-adjoint if the spectrum of a  self-adjoint Hamiltonian  is
bounded from below. As a consequence, the time-energy uncertainty
relation cannot be deduced from the same kinematical point of view
as the position-momentum uncertainty relation.
 Nevertheless, the search for various time operators and the analysis of their
 self-adjointness and associated time-energy uncertainty relations
 have been the subject of a number of papers [4]. The general consensus
 seems to be that no such self-adjoint operator exists.

 Recently, the validity of  Pauli's objections has been
 critically evaluated [5], with the conclusion that there is no a
 priori reason to exclude the existence of  self-adjoint time
 operators for semibounded Hamiltonians.

In this work we consider the  problem of constructing a
self-adjoint  time operator for a singular harmonic oscillator.

\section{The problem}
The singular harmonic oscillator of the Calogero-Sutherland type
[6] is described by the Hamiltonian
\begin{equation}
H_{CS}   \equiv 2\omega K_3  = \omega ^2K + H,
\end{equation}
where $K = x^2/2$ and
\begin{equation}
H = \frac{1}{2} ( p^2 + \frac{g}{x^2}), \,\, g > 0.
\end{equation}
is the Calogero-Moser [7] scale invariant Hamiltonian. We have
identified $H_{CS}$ Hamiltonian with the compact generator, $K_3$,
of the SU(1,1) group, which is the  dynamical group of this
problem. Two other generators of  $SU(1,1)$  are
\begin{eqnarray}
K_1 &  =  & \frac{1}{2}(\omega K -
\frac{1}{\omega}H),\nonumber\\
K_2 & = & D,
\end{eqnarray}
where $D = -(xp + px)/4$ is the scale operator.

The  group generators $K_3$ and $K_{\pm} = K_1 \pm iK_2$ satisfy
the standard commutation relations of the su(1,1) algebra:
\begin{equation}
[K_3 , K_{\pm}] = \pm K_{\pm},  \,\,\,\, [K_-, K_+ ] = 2K_3.
\end{equation}
Our objective  is to construct an operator $\hat{T}$ in terms of
the generators $K_3,K_{\pm}$ that is  conjugate to the Hamiltonian
$H_{CS}$ and satisfies $[H_{CS},\hat{T}] = i$.

Let us denote by $|n,k> , n = 0,1,2,...$ the  complete orthonormal
basis states, which  diagonalize the compact generator $K_3 =
H_{CS}/2\omega $ [8].  The Bargman index  $k = \frac{1}{2}(1 +
\sqrt{g + \frac{1}{4}})$ is related to the eigenvalue $k(k-1)= g$
of the quadratic  Casimir operator $\hat{C}_2 = K_{3}^2 - K_{1}^2
- K_{2}^2$ of the $SU(1,1)$ group. These states are obtained from
$|0,k>$ by n-fold application of $K_{+}$:
\begin{eqnarray}
|n,k> & = & \sqrt{\frac{\Gamma (2k)}{\Gamma (2k+n)n!}} (K_{+})^n
|0,k>,
 \nonumber\\ K_{-}|0,k> & = & 0,\\
K_3 |n,k> & = & (n+k)|n,k>,\,\, n = 0,1,2,... \nonumber
\end{eqnarray}

In the space spanned by the eigenstates of the generator $K_3$, we
immediately encounter the problem. The matrix elements of
$[H_{CS},\hat{T}]$ in the  basis $|n,k>$,
\begin{equation}
<n,k|[H_{CS},\hat{T}]|m,k> = 2\omega (n - m)<n,k|\hat{T}|m,k>
\end{equation}
vanish for $n = m$. This  implies $[H_{CS},\hat{T}] \neq i$ if
$<n,k|\hat{T}|m,k> \neq 0$. This relation is correct only if the
operation by $\hat{T}$ on a state $|n,k>$ is of the form
\begin{equation}
\hat{T}|n,k> = \sum _m t_{mn}|m,k>.
\end{equation}
However, $\hat{T}$ does not have that property if it is conjugate
to $H_{CS}$. In the next Section, we shall study
$[H_{CS},\hat{T}]$ commutator in the time-variable representation,
i.e., in the representation in which $\hat{T}$ is diagonal.

\section{Construction of $\hat{T}$}

Let us observe that from $[K_3, K_-^n] = - n K_-^n$, we can make a
simple ansatz that $\hat{T}$ is some power series function of
$K_-$ and $K_+$ such that
\begin{equation}
[K_3, F(K_{\pm})] = \pm K_{\pm}F'(K_{\pm}) = \frac{i}{4\omega }.
\end{equation}
A possible solution is
\begin{equation}
\hat{T} = \frac{1}{4i\omega }(lnK_- \, - \, lnK_+),
\end{equation}
which is easily represented in the coherent state representation
of the operator $K_-$.

 States which diagonalize the operator $K_-$
\begin{equation}
K_-|z,k>  =  z|z,k>,
\end{equation}
where $z$ is an arbitrary complex number, are known as the
Barut-Girardello (BG) coherent states [9], [10]. The expansion of
these states over the orthonormal basis  $|n,k>$ is
\begin{equation}
|z,k\rangle = \frac{z^{k-1/2}}{\sqrt{I_{2k-1}(2|z|)}} \sum_{n =
0}^{\infty} \frac{z^{n}}{\sqrt{n!\Gamma(2k+n)}} |n,k\rangle ,
\end{equation}
where $I_\nu(z)$ is the  modified Bessel function of the first
kind. The above BG states are normalized to unity, they resolve
the identity operator, but  are not mutually orthogonal
\begin{equation}
<z_{1},k|z_{2},k >= I_{2k-1}(2\sqrt{z_{1}^{\ast}z_{2}})[
I_{2k-1}(2|z_{1}|) I_{2k-1}(2|z_{2}|)]^{-1/2}.
\end{equation}
Due to this property any quantum  state $| \psi >$ can be
represented by the analytic function
\begin{equation}
f_{\psi}(z) = \sqrt{I_{2k-1}(2|z|)}(z^{1/2- k}) <
k,z^{\ast}|\psi>.
\end{equation}
The operators ${K}_{\pm}$ and ${K}_{3}$ act in the Hilbert space
of analytic functions $f_{\psi}(z)$ as linear differential
operators

\begin{equation}
{K}_{+} =  z \ , \quad {K}_{-} = 2k \frac{d}{dz} + z
\frac{d^{2}}{dz^{2}} \ , \quad {K}_{3} = k + z \frac{d}{dz} \ .
\end{equation}
In terms of  BG coherent states, the time operator $\hat{T}$ is
\begin{equation}
\hat{T} = (4\pi i)^{-1}\int d\mu(z,k)ln(\frac{z}{z*})|z,k><z,k|,
\end{equation}
where $d\mu(z,k) = 2K_{2k-1}(2|z|)I_{2k-1}(2|z|)d^2z/\pi $.

\section{Discussion}

BG coherent states can also be written as an exponential operator
acting on the vacuum state of $K_-$,
\begin{equation}
|z,k> = e^{zK_+ (K_3 +k)^{-1}}|0,k>.
\end{equation}
In deriving this expression, we have used an operator identity\\
\begin{equation}
[K_+ (K_3 +k)^{-1}]^n = K_+^{n} \frac{\Gamma (K_3 +k)}{\Gamma (K_3
+k +n)}.
\end{equation}
Note also that the operator $K_+ (K_3 +k)^{-1}$ is canonical to
$K_-$:
\begin{equation}
[K_-,K_+ (K_3 + k)^{-1}] = 1.
\end{equation}

It is  easy to see that $H = \omega (K_3 - K_1)$ is  related to
$K_-$:
\begin{equation}
e^{-\omega K} H e^{\omega K} = -2\omega K_-.
\end{equation}
Therefore, the  energy eigenstates of $H|E> = E|E>$ are
proportional to the BG coherent states [9,11] if  $z = -E/2\omega
$:
\begin{equation}
|E> = e^{\omega K}|- \frac{E}{2\omega},k>.
\end{equation}
Note also, that the state $<x|E>$, in the limit $E\rightarrow 0$,
is not normalizable, since lim$_{E\rightarrow 0}<x|E> =
<x|e^{\omega K}|0,k> \propto \omega^k x^{2k-1/2}$. \\The
difficulty arises from the oscillating behavior of $<x|E>$ at
large distances [12].

Finally, we consider an explicit  construction of  time operator
for  $H_{CS}$ using the method developed in [13,14,15]. We first
observe that there exists a singular similarity transformation
between $H_{CS}$ and the Hamiltonian of the ordinary
 harmonic oscillator, $H_h =  H_{CS}(g=0)$ :
\begin{eqnarray}
H_{CS} S & = &  S H_h, \nonumber\\
S & = & e^{-K_-} e^{K_{-}^{0}},
\end{eqnarray}
where $K_{-}^{0} = K_{-}(g=0)$. The time operator for $H_h$ was
 constructed and discussed  in [14, 15, 16]. Its construction is
 simple if we observe that the Casimir operator with $k = 3/4$ can be
 used to express the operator $K$ in the form
\begin{eqnarray}
K & = & T_0 H_0 T_0 + \frac{1}{16H_0} \nonumber \\
& = & QH_0Q - \frac{i}{2}Q, \\
Q & = & - T_0 + \frac{i}{4H_0}.\nonumber
\end{eqnarray}
where
\begin{equation}
T_0  =  - \frac{1}{2}(x\frac{1}{p} + \frac{1}{p}x)
\end{equation}
is the time-of-arrival operator of Aharonov and Bohm [16], and
$H_0 = p^2/2$. Then the Hermitian operator
\begin{equation}
T_h = \frac{1}{2}(T_h(Q) + T_h^{\dagger }(Q))
\end{equation}
satisfies $[H_h,T_h] = i$, where
\begin{equation}
T_h(Q) = \frac{1}{\omega }arctg (\omega Q).
\end{equation}
It is now easy to see that the time operator for the Hamiltonian
$H_{CS}$ is
\begin{equation}
T_{CS} = S T_h S^{-1}, \;\; S^{-1} \neq S^{\dagger }.
\end{equation}
Note that in this construction $T_{CS} \neq T_{CS}^{\dagger }$.
Formally, in the limit $\omega \rightarrow 0 $ we obtain the time
operator for  the scale invariant Hamiltonian $H$ [17].

\section{Conclusions}

In conclusion, we have presented an algebraic method of
constructing  Hermitian operators conjugate to a Hamiltonian with
$SU(1,1)$ dynamical symmetry. In terms of  generators of
$SU(1,1)$, the time operator for a singular harmonic oscillator is
constructed explicitly, and shown that it can be  related to the
time-of-arrival operator, $T_0$ of Aharonov and Bohm. The question
whether  time operators thus constructed are self-adjoint
operators in Hilbert space requires a careful examination of their
spectra and eigenfunctions. The time-energy commutation relation
is  studied  in the energy and the time domains. The eigenvalue
problem of the operator $T_{CS}$ can be solved in the time domain
using BG coherent states. It is not self-adjoint and its
eigenfunctions are not orthogonal. Therefore, the problem of
finding self-adjoint  $T_h$ and $T_{CS}$  is still open [5,18].
Let us also mention  that the problem of time-operator   for a
repulsive singular potential of the Calogero-Moser type [7]  is
interesting for several reasons. It is scale invariant and  has
the full conformal group $SO(2,1)$ as a dynamical symmetry group
[12] with the generators $H, D$ and $K$. The spectrum of $H$, for
$g > 0$ is positive, continuous, and bounded from below, but with
a non-normalizable ground state. $H$ can  be easily extended to
the well-known one-dimensional N-body problem of Calogero-Moser
[7].  Recently, it has been observed that the dynamics of  scalar
particles near the horizon of a black hole is also associated with
this Hamiltonian [17,19,20,21].

It is important to point out here that the solution of the formal
equation $[H,\hat{T}]=i$ is not unique. Any $\hat{T}' = \hat{T} +
\phi (H)$, with arbitrary $\phi$, satisfies the same canonical
commutation relation.

{\bf Acknowledgment}

This work was supported by the Ministry of Science and Technology
of the Republic of Croatia under Contract No.0098004.

\end{document}